\documentclass[conference,letterpaper]{IEEEtran}
\usepackage{booktabs}

\newcommand{\citep}{\cite}
\newcommand{\citet}{\cite}

\usepackage{times}

\usepackage[linesnumbered, ruled, vlined]{algorithm2e}      
\usepackage{float}
\usepackage{stackengine}
\usepackage{stmaryrd}
\usepackage{dsfont}
\usepackage[T1]{fontenc}
\usepackage{url}

\usepackage[cmex10]{amsmath}  
\usepackage{amsfonts}
\usepackage{mathrsfs}
\usepackage{tikz}
\usepackage[utf8]{inputenc}

\usepackage{epstopdf}
\usepackage{mathtools}
\usepackage{graphicx}
\usepackage{bbm}
\usepackage{enumerate}
\usepackage{epsf,verbatim,amssymb,array,cite,multicol,multirow}  
\usepackage{psfrag,bm,xspace}
\usepackage{hhline}
\usepackage{xstring}
\usepackage{ifthen}

\usepackage{xparse}
\usepackage{physics}

\ifdefined \theorem 
\else
  \newtheorem{theorem}{Theorem}
\fi
\newtheorem{lem}{Lemma}

\ifdefined \proposition 
\else
\newtheorem{proposition}{Proposition}
\fi

\ifdefined \definition 
\else
\newtheorem{definition}{Definition}
\fi

\ifdefined \example 
\else
\newtheorem{example}{Example}
\fi

\ifdefined \remark 
\else
\newtheorem{remark}{Remark}
\fi

\usepackage{xcolor}
\definecolor{darkblue}{rgb}{0.1,0.1,0.8}
\definecolor{brickred}{rgb}{0.8, 0.25, 0.33}
\definecolor{britishracinggreen}{rgb}{0.0, 0.26, 0.15}
\definecolor{calpolypomonagreen}{rgb}{0.12, 0.3, 0.17}
\definecolor{ao(english)}{rgb}{0.0, 0.5, 0.0}
	\definecolor{cadmiumgreen}{rgb}{0.0, 0.42, 0.24}
\definecolor{burgundy}{rgb}{0.5, 0.0, 0.13}
\usepackage{etoolbox}

\providetoggle{Blue_revision}
\settoggle{Blue_revision}{true}

\providetoggle{Track}
\settoggle{Track}{true}
\newcommand{\addv}[3]{%
	\iftoggle{Track}{%
    	\IfEqCase{#1}{%
       	 	{a}{\ifthenelse{\equal{#2}{ON}}{{\color{cadmiumgreen}#3}}{#3}}%
        	{b}{\ifthenelse{\equal{#2}{ON}}{{\color{brickred}#3}}{#3}}%
       		{c}{\ifthenelse{\equal{#2}{ON}}{{\color{burgundy}#3}}{#3}}%
    	}[\PackageError{tree}{Undefined option to tree: #1}{}]%
	}{#3}%
}

 \usepackage{hyperref}
\usepackage{xcolor}
\definecolor{DarkGreen}{rgb}{0,0.6,0}

\hypersetup{
colorlinks=true, %
  pdfstartview={FitH},
    linkcolor=red,
    citecolor=blue,
    urlcolor={blue!80!black}
}

\usepackage{graphicx}

\newcounter{relctr} 
\everydisplay\expandafter{\the\everydisplay\setcounter{relctr}{0}} 

\AtBeginDocument{} 

\global\long\def\RR{\mathbb{R}}

\global\long\def\NN{\mathbb{N}}

\global\long\def\EE{\mathbb{E}}
\global\long\def\PP{\mathbb{P}}

\global\long\def\11{\mathbbm{1}}

\newcommand{\bfx}{\mathbf{x}}

\newcommand{\bfz}{\mathbf{z}}

\global\long\def\+{\oplus}

\newcommand\pmm{\{-1,1\}}

\newcommand{\prob}[1]{\PP\Big\{  #1 \Big\} }
\def\<{\langle}
\def\>{\rangle}

\ifdefined \var 
  \renewcommand{\var}{\mathsf{var}}
\else
  \newcommand{\var}{\mathsf{var}}
\fi

\ifdefined \abs \else
 \newcommand{\abs}[1]{\lvert#1\rvert}
\fi

\ifdefined \norm \else
 \newcommand{\norm}[1]{\lVert#1\rVert}
\fi

\ifdefined \set 
  \renewcommand{\set}[1]{\left\{#1\right\}}
\else
  \newcommand{\set}[1]{\left\{#1\right\}}
\fi

\newcommand*{\medcup}{\mathbin{\scalebox{1}{\ensuremath{\bigcup}}}}%

\DeclareMathOperator*{\argmin}{arg\,min}

\usepackage{stackengine}
\def\deq{\mathrel{\ensurestackMath{\stackon[1pt]{=}{\scriptstyle\Delta}}}}

\newcommand{\optfont}[1]{\mathsf{#1}}
\DeclareMathOperator*{\tensor}{\otimes}

\providecommand{\tr}{tr}

\ifdefined \Tr 
  \renewcommand{\Tr}[1]{\tr \Big\{#1\Big\}}
\else
  \newcommand{\Tr}[1]{\tr \Big\{#1\Big\}}
\fi

 \def\id{I_d}
\global\long\def\bigtensor{\bigotimes}

\newcommand{\opt}{\optfont{opt}}
\def\qmstate{\sigma^M_{Y\hat{Y}}}

\def\define{:=~}
\def\ulineb{\underline{b}}
\def\hatb{\hat{b}}
\def\ulinehatb{\underline{\hat{b}}}

\def\QERMPOVM{\mathcal{L}_{\text{QERM}}^{\mathcal{C}}}

\def\MLoss{\mathcal{L}_M}
\usepackage[nolist,nohyperlinks]{acronym}

\newacro{ptp}[PtP]{Point-to-Point}
\newacro{iid}[i.i.d.]{independent and identically distributed} 
\newacro{IID}[i.i.d.]{independent and identically distributed} 
\newacro{UFFS}[UFFS]{Unsupervised Fourier Feature Selection}
\newacro{SFFS}[SFFS]{Supervised Fourier Feature Selection}
\newacro{LS}[LS]{Laplacian Score}
\newacro{MAE}[MAE]{mean absolute error}
\newacro{MSE}[MSE]{mean square error}
\newacro{PAC}[PAC]{probably approximately correct}
\newacro{VC}[VC]{Vapnik–Chervonenkis}
\newacro{ERM}[ERM]{Empirical Risk Minimization}
\newacro{SVM}[SVM]{support-vector machine}
\newacro{POVM}[POVM]{positive operator-valued measure}
\settoggle{Track}{false}
\settoggle{Blue_revision}{false}
\newcommand{\addva}[1]{\addv{a}{off}{#1}}

\newcommand{\addvd}[1]{\addv{a}{off}{#1}}
\newcommand{\addvf}[1]{\addv{a}{off}{#1}}

\newcommand{\addvg}[1]{\addv{b}{ON}{#1}}

\begin{document}
\title{A Theoretical Framework for Learning from Quantum Data}

\author{
\IEEEauthorblockN{Mohsen Heidari}
\IEEEauthorblockA{Purdue University\\
mheidari@purdue.edu}
\and
\IEEEauthorblockN{Arun Padakandla}
\IEEEauthorblockA{University of Tennessee\\
arunpr@utk.edu}
\and
\IEEEauthorblockN{Wojciech Szpankowski}
\IEEEauthorblockA{
Purdue University\\
szpan@purdue.edu}
}
\maketitle

%

\begin{abstract}
Over decades traditional information theory of source and channel coding advances toward learning and effective extraction
of information from data. We propose to go one step further and offer a theoretical foundation for learning classical patterns
from {\it quantum data}. However, there are several roadblocks to lay the groundwork for such a generalization.
First, classical data must be replaced by a density operator over a Hilbert space. Hence, deviated from problems such as \textit{state tomography}, our samples are i.i.d density operators.  The second challenge is even more profound since we must realize that our only interaction with a quantum state
is through a measurement which -- due to no-cloning quantum postulate -- loses information after measuring it. With this in mind,
we present a quantum counterpart of the well-known \ac{PAC} framework. Based on that, we propose a quantum analogous of the \ac{ERM} algorithm for learning measurement hypothesis classes. Then, we establish upper bounds on the quantum sample complexity quantum concept classes. 
\end{abstract}

\section{Introduction}
\label{Sec:Introduction}
Over the past few decades, we have been mastering the ability to \textit{learn} from data to perform many tasks such as classification, statistical inference, and pattern recognition. Recent achievements in quantum information processing to collect, store, and process quantum systems endow us with a more powerful ability: learning from quantum data.

 As research in quantum information theory suggests, fundamental concepts in classical settings admit multiple quantum counterparts. For example, the task of communicating data over quantum channels leads to multiple notions of \textit{capacity}\cite{Wilde2013}. The task of ``learning" from ``quantum data" is not an exception. Recently, researchers have been developing different learning frameworks \cite{Aaronson2007,Cramer2010,Barnett2009,Bshouty1998,Badescu2019}. 
 
From the perspective of quantum statistical learning theory, which is the view of this work, the learning models can be grouped into two main categories. 
The first category, refered to as  \textit{state tomography} or \textit{state discrimination},  the objective is to find an approximate description of an unknown quantum state or distinguish it from another state using \textit{measurements} on multiple copies of the state \cite{ODonnell2016,ODonnell2017,Haah2016}. \addvd{A survey on this topic is provided in \cite{Montanaro2016}. An operational view of learning quantum states is introduced by \cite{Aaronson2007}. Another related work in this line is \cite{Cheng2015} where the objective is to learn an unknown measurement $E$ from samples of the form $\set{(\rho_i, \tr{E\rho_i})}_{i=1}^n$, where $\rho_i$'s are \ac{iid} random quantum states.} Quantum state classification in this model then studied under various restrictions on the states (e.g., pure, mixed) \cite{Gambs2008,Guta2010}.
In the second group of works, which is referred to as the \textit{quantum oracle model}, we measure identical copies of a superposition state to solve a classical learning problem \cite{Bshouty1998,Arunachalam2017}. Learning using this method has been explored in several works such as \citep{Arunachalam2017,Kanade2018,Bernstein1997,Servedio2004} and analogous of the well-known \textit{agnostic} \ac{PAC} framework was introduced in \citep{Arunachalam2018}.

The main departure point of this article from the mentioned models stems from the fact that samples are not identical copies of each other; rather, they are \ac{iid} quantum states. \addva{Further, we are not required to learn the states. Rather we need only to learn a classical attribute to such states. That is, we have an \textit{ensemble} of quantum states, and associated with each state, we have a classical attribute/label. Alternatively, one can think of a quantum system that is measured by an unknown measurement (nature's measurement). We have access to the post-measurement states as well as the classical outcomes. The objective is to learn this measurement.}
\addvf{  Applications of this model has been studied under various settings \cite{Schuld2019,Mitarai2018,Schuld2020} such as classification of entangled and separable quantum states  \cite{Gao2018,Ma2018} and integrated quantum photonics \cite{Kudyshev2020}.} That said, we propose a different model for learning from quantum data. As a prototype, consider the following problem:

Suppose a physical device randomly emits a sequence of quantum states (e.g., photons), say $\rho_1, \rho_2, ...$. Associating to each state is a classical attribute $y_i\in \mathcal{Y}$,  such as ``red" or ``blue" as its color. The probability distribution of the states and the underlying law governing their classical attribute are unknown. However, we know that the states belong to a family of parametrized quantum systems.    We seek a procedure that, given a number of training quantum states with their labels, learns the device's coloring/labeling law to predict the label of a new quantum state from this device. 


Our problem formulation is motivated by the original/early questions that led to the theory of statistical learning. Suppose a computing device is provided with $m$ training samples $(x_{i},y_{i}) \in \mathcal{X}\times \mathcal{Y} : 1 \leq i \leq m$, can it learn the probabilistic/functional relationship between the label $y \in \mathcal{Y}$ and the features $x \in \mathcal{X}$. More specifically, under what conditions can an algorithm pick out a function from its library (hypothesis class) that best approximates the probabilistic/functional relationship? The pursuit of an answer to this question led to the elegant theory of PAC learning, \ac{VC} dimension, Rademacher complexity and such. As we describe in the sequel, our work formulates this very question in a quantum setup and we provide an initial set of our findings.

As our first contribution, we propose a quantum counterpart of the \ac{PAC} learning framework as developed by \cite{Kearns1994,Valiant1984}. In our model, the samples are pairs $(\rho_i, y_i)$, where $\rho_i$'s are density operators on a Hilbert space $H_X$ and $y_i\in \mathcal{Y}$ are the classical labels.  What we therefore seek is a measurement that will label a quantum state correctly. Hence, the predictors are measurements modeled as \ac{POVM}.   Analogous to the standard PAC, our quantum algorithm has a library of \ac{POVM}s modeling the concept class of candidate predictors. By fixing a loss operator, we are lead to the analogous fundamental question of PAC learning: What is the quantum sample complexity for learning a measurement class?

To answer this question, we propose the quantum analogous of ERM algorithm and provide a bound on the quantum sample complexity. We will show that our model subsumes the classical PAC framework under some orthogonality conditions. Further, our sample complexity bounds match with classical ones. As a result, we conclude that the task of learning from quantum states is harder than classical. In other words, quantum sample complexity is not smaller than the classical sample complexity. We further show that the quantum sample complexity of a quantum concept class depends not only on its size but on a fundamental property called \textit{compatibility} of the measurements in the class \cite{Holevo2012}. Such intrinsic quantum nature of the problem precludes a straightforward use of already developed complexity measures such as VC dimension, covering number and fat-shattering dimension \cite{Kearns1990}, and Rademacher complexity from statistical learning theory \cite{Shalev2014}.

As a careful reader will recognize, this learning framework hides several complexities.  In what follows, we briefly highlight some of its challenges and differences from previous models.



First, our only interaction with a quantum state is through measurement. This necessitates the learning algorithm to be implemented via a \textit{quantum measurement} with possible classical post-processing. Hence, abiding axioms of quantum mechanics, we can process the training samples only once, as they collapse after the measurement. This is a challenge; because, unlike the mentioned models,  we do not have access to identical copies of the training samples. This difficulty is exacerbated as the \textit{no-cloning} principle prohibits making new copies from the states at hand. 

The second challenge arises from the \textit{uncertainty principle}. Usually, a learning algorithm needs to estimate multiple parameters via different measurements on the samples (e.g., empirical loss of different predictors). Ideally, we would like to combine these measurements and use one set of samples for all estimations. However, such measurements might not be \textit{compatible} and hence, if we combine them, the estimations' accuracy can drop significantly \cite{Holevo2012}. Motivated by the notion of \textit{unbiased measurements}  \cite{Lawrence2002,Bandyopadhyay2001}, we  propose  \textit{compatibility} covering in Section \ref{sec:main}.

Third, the training states are not completely distinguishable as they are not orthogonal. Hence, the amount of information we can extract from the samples is limited by the amount of their overlaps.

This paper is organized as follows:  In Section \ref{sec:model}, we formally describe the elements of our model and define a new quantum analogous of PAC. Then, in Section \ref{subsec:classical subsume} we argue that classical learning is subsumed under this model.  In section \ref{sec:main} we elaborate on the compatibility issue and propose our sample complexity bound. Lastly, in Section \ref{subsec:QERM} we propose Quantum ERM (QERM) to prove our results.

\noindent\textbf{Preliminaries:}
Quantum states, as usual, are density operators that are linear,  self-adjoint, unit-trace, and positive semi-definite. We denote by $\mathcal{D}(H)$ the set of all density operators on $H$. 
Any quantum measurement in this paper is modeled by a \ac{POVM}. We denote a POVM as  $\mathcal{M}:=\{M_v, v\in\mathcal{V}\}$, where $\mathcal{V}\subset \RR$ is the (finite) set of possible outcomes. Operators of the measurement satisfy the following conditions: $M_v=M_v^\dagger \geq 0,  \sum_v M_v =I,$
  where $I$ is the identity operator. \addvd{For short-hand, we use $[n]$ to denote the set $\set{1,2,...,n}$ for any $n\in \NN$.}  

\section{The proposed Quantum Learning Model}\label{sec:model}
In this section, we formally propose our learning model. We discuss the differences between this model and the standard PAC framework. Also, we show that the classical learning framework is subsumed under our model.

Similar to the PAC framework, our model consists of multiple components, which are defined in the following. Let $\mathcal{X}$ be a finite set.  The feature set is a collection of fixed density operators $\rho_x, x \in \mathcal{X}$, acting on a fixed Hilbert space $H_X$.  The set of possible classical labels is a finite set $\mathcal{Y}$. For example, in the binary classification of qubits, $H_X$ is a two-dimensional Hilbert space and $\mathcal{Y}=\{0,1\}$. 

For  compactness, we consider an auxiliary quantum register (pure state) for storing the classical labels. Let $H_Y$ denote the Hilbert space of the labels created as $H_Y =\text{span}\set{\ket{y}: y\in \mathcal{Y}}$.
 With this notation, $\rho_x$ together with its label $y$ are represented by the bipartite quantum state $\rho_x\tensor \ketbra{y}$. Hence, the feature-label set is given by $\set{\rho_x\tensor \ketbra{y}: x\in \mathcal{X}, y\in \mathcal{Y}}$.

Consider an unknown, but fixed, probability distribution $D$ on $\mathcal{X}\times \mathcal{Y}$.  As the training set, we are given $n$ \ac{iid} samples $\rho_{x_i}\tensor \ketbra{y_i}, i\in [n]$, where $(x_i,y_i)$ are drawn from $D$. 
With this setup, the training samples are represented by the tensor product state $S_n = \bigtensor_{i=1}^n \big(\rho_i \tensor \ketbra{y_i} \big)$. Further, the average density operator of each sample is $\rho_{XY}=\sum_{x,y} D(x,y) \rho_x\tensor \ketbra{y}$. 

We seek a procedure that, given the training samples, construct a predictor for the task of classification (statistical inference). The predictor is given the only feature state $\rho_x$ and is tasked to produce a label.  Since the features are quantum states and the labels are classical, the predictors are quantum measurements.
That said, a predictor is a POVM $\mathcal{M}:=\set{M_{{y}}: {y}\in {\mathcal{Y}}}$ acting on the $X$-system only. To test a predictor $\mathcal{M}$, a new sample is drawn according to $D$. If $\rho_x\tensor \ketbra{y}$ is the realization of the test sample, then without revealing $y$, we measure $\rho_x$ with $\mathcal{M}$. The outcome of this predictor is $\hat{y}$ with probability $\tr{M_{\hat{y}}\rho_x}, \hat{y}\in \mathcal{Y}$. Note that this is different from the classical settings, where the output of the predictor is a deterministic function of the samples. 
Since our labels are essentially stored in classical registers, we employ a conventional loss function to measure the accuracy of the predicted label. Thus, by  $\ell : \mathcal{Y}\times {\mathcal{Y}}\mapsto [0,1]$ we denote the (normalized) loss  function. Therefore, the true risk of a predictor $\mathcal{M}$ with respect to the underlying sample's distribution $D$ is  
 \begin{align*}
 L_D(\mathcal{M})  \deq  \sum_{(x,y,\hat{y})\in \mathcal{X}\times \mathcal{Y}\times {\mathcal{Y}}} D(x,y) ~\ell(y,\hat{y})~ \tr{M_{\hat{y}}\rho_x}.
 \end{align*}

The concept class in our model is a collection $\mathcal{C}$ of predictors and its minimum loss is denoted by $\opt_{C} \deq \inf_{\mathcal{M}\in \mathcal{C}} L_D(\mathcal{M}).$
Before describing the rest of the model, let us present the following example. 
\begin{example}
 \label{Ex:ElectronSpin1}
 Consider electrons with spin pointing in a direction, represented by a $3-$dim unit vector in the \textit{Bloch sphere}. Let finite set $\mathcal{X}=\{ (\theta_{i},\phi_{j}) = (\frac{i\pi}{20},\frac{j2\pi}{20}):0\leq i,j\leq 19\}$ represent the possible spin axis directions. We have two labels in $\mathcal{Y} = \{\mbox{blue},\mbox{red}\}$. Nature decides to label an electron `blue' if the axis of its spin is orthonormal to a specific orthant. Otherwise, the electron is labeled `red'. For this, she chooses a specific orthant $\mathcal{O}$. This establishes a relationship - $p_{Y|X}$ - between the elements $(x,y) \in \mathcal{X} \times \mathcal{Y}$. Going further, she chooses a distribution $p_{X}$, samples $X$ wrt this distribution, endows an electron with the corresponding spin, and hands only the electron to us. Our predictor is aware of $\mathcal{X}$, its association with the spin directions, i.e., the mapping $x \rightarrow \rho_{x}$, and $\mathcal{Y}$. Oblivious to both nature's decision and the orthant, but possessing the prepared electron, a predictor's task is to unravel the label. The predictor is a measurement with two outcomes, `blue' and `red'. An optimal predictor will be able to distinguish whether the axis of an electron's spin is orthonormal to $\mathcal{O}$ or otherwise. 
\end{example}

\noindent\textbf{Learning Algorithm as a Quantum Measurement:}
A quantum learning algorithm is a process that, with the training samples as the input, selects a predictor from the concept class.\footnote{Our focus is on \textit{proper} algorithms. Generally, we allow the selected predictor to be outside of the concept class.} This process is modeled as a quantum measurement on the joint space of all training the samples, i.e., $H_{XY}^{\tensor n}$. The outcome of this measurement is a  classical number as the index of the selected predictor in the concept class.  

{\begin{definition} \label{def:quantum algorithm}
Let $H_{XY}$ be the feature-label Hilbert space. Also let $\mathcal{C}$ be the concept class whose members are indexed by a set $\mathcal{J}$. Then, a  (\textit{proper}) quantum learning algorithm is a sequence of POVMs $\mathcal{A}_n:=\set{A_{n,j}: j\in \mathcal{J}}, n\in \NN$, acting on $H_{XY}^{\tensor n}$, the space of $n$ samples, and with outcomes in $\mathcal{J}$. 
\end{definition} 
 Unlike the classical settings, even if the samples are fixed, the algorithm's output is a random variable on $\mathcal{J}$.  That said, we can write $M_J\in \mathcal{C}$ as the selected predictor with $J$ being a random variable on $\mathcal{J}$. 
 With all the components described, we are ready to define the quantum version of PAC learnability. 
 
\begin{definition}[QPAC] \label{def:quantum PAC}
Given a concept class $\mathcal{C}$, an algorithm $\mathcal{A}_n, n\in \NN$  QPAC learns $\mathcal{C}$,  if there exists a function $n_{\mathcal{C}}: (0,1)^2\mapsto \NN$ such that for every $\epsilon, \delta \in (0,1)$ and all $n\geq n_{\mathcal{C}}(\epsilon,\delta)$  
\begin{align*}
\sup_D \sum_{j\in \mathcal{J}} \tr{A_{n,j} \rho_{XY}^{\tensor n}} \11\set{L_D(\mathcal{M}_j)> \opt_{\mathcal{C}}+\epsilon}\leq \delta,
\end{align*}
where $\rho_{XY}$ is the average density operator of the samples with respect to $D$ and  $\mathcal{M}_j\in \mathcal{C}$ is the $j$th predictor in the class. 
\end{definition}} 
 


Our goal is to characterize concept classes that are learnable and quantify their sample complexity. Before that, let us discuss the connection to the classical PAC.

\subsection{Classical PAC learning is a special case}\label{subsec:classical subsume}
We argue that the proposed formulation subsumes the classical PAC learning framework. 
\begin{theorem}
For a classical PAC learning model with  feature-label set $\mathcal{X}\times \mathcal{Y}$, hypothesis class $\mathcal{H}$, loss function $l:\mathcal{Y}\times \mathcal{Y} \mapsto [0,1]$, and algorithm $A$, there exist a corresponding element in the quantum learning model such that $A$ is a PAC learning algorithm with respect to the classical model if and only if its quantum counterpart is a QPAC learning algorithm under the quantum model. 
\end{theorem}
\begin{IEEEproof}[Proof idea]
We set $\rho_x=\ketbra{x}, \forall x\in\mathcal{X}$, where $\ket{x}$'s are pure orthogonal states. As a result the feature-label density operators are $\ketbra{x}\tensor \ketbra{y}, x\in \mathcal{X}, y\in\mathcal{Y}$. 
As for the quantum hypothesis class, for any $f\in \mathcal{H}$ define the POVM $\mathcal{M}_f =\set{M_y^f: y\in \mathcal{Y}}$ where  $M_y^f \deq \sum_{x: f(x)=y} \ketbra{x}$. Then, our hypothesis class $\mathcal{C}$ is the collection of such POVMs $\mathcal{M}_f, f\in \mathcal{H}$. It is not difficult to see that the risk of any predictor $\mathcal{M}_f$ equals
 $L_D(\mathcal{M}_f)=\EE_D[l(Y,f(X))]$ which is the classical risk of $f$. Further, since the states are completely distinguishable, one can show that any classical learning algorithm can be implemented by a quantum algorithm.  As a result, of these arguments, we can show that Definition \ref{def:quantum PAC} reduces to the standard PAC definition and that the classical sample complexity matches with quantum samples complexity. 
\end{IEEEproof}

Note that in the setting of the above result, beyond possible computational advantages, the quantum learning does not benefit statistically. Hence, in this case, the quantum sample complexity matches the classical one. However, this might not be the case when the hypothesis class is classical, but $\rho_x$'s are not orthogonal. Similarly, in quantum source coding, when the states are not orthogonal, we get an advantage in compression rates \citep{Datta2013,Schumacher1995}. 

\section{Quantum PAC Learning Results }\label{sec:main}
In this section, we present our main results, which is a bound on quantum sample complexity. As discussed in the introduction, our bounds depend on the \textit{compatibility} structure of the predictors in the concept class. To present our results, we need to elaborate on the notion of \textit{compatibility}. 
\addvd{The predictors in this paper are assumed to be \textit{sharp} measurements. Thus, from Theorem 2.13 of \cite{Holevo2012} the definition of compatibility is reduced to the following. }

\begin{definition} \label{def:compatible}
A collection of sharp measurements $\mathcal{M}^j=\set{M^j_y: y\in \mathcal{Y}}, j =1,2, ..., k$, are \textit{compatible} if their operators mutually commute, that is $M^j_y M^\ell_{\tilde{y}}= M^\ell_{\tilde{y}}M^j_y$ for all $j,\ell\in [k]$ and all $y,\tilde{y}\in \mathcal{Y}$.
\end{definition}

Consequently, if $\mathcal{C}$ is a compatible concept class, then there exists a basis on which all the predictors are diagonalized. If $\mathcal{C}$ is a general concept class. Then, we group its members into  compatible subclasses. 

\begin{definition} \label{def:compatible cover}
Given  a collection of observables $\mathcal{C}$, a compatibility partitioning is a family of distinct subsets $\mathcal{C}_1, \mathcal{C}_2, ..., \mathcal{C}_m$ of $\mathcal{C}$ such that $\mathcal{C}=\medcup_r \mathcal{C}_r$ and that the observables inside each $\mathcal{C}_r$ are compatible internally with each other. 
\end{definition} 

 Note that there always exists a compatibility partitioning as the single element subsets of $\mathcal{C}$ form a valid covering. Further, note that the compatibility structure is an inherent property of the concept class which is independent
of the samples. 

Now with the above definitions, we are ready to present our main result in the following theorem.

\begin{theorem}\label{thm:QERM}
Any finite hypothesis class  $\mathcal{C}$ is agnostic QPAC learnable with quantum sample complexity bounded as
\begin{align*}
n_{\mathcal{C}}(\epsilon, \delta) \leq  \min_{\mathcal{C}_r \text{Comp. partition}} \sum_{r=1}^m \Big\lceil{\frac{8}{\epsilon^2}\log\frac{2m|\mathcal{C}_r|}{\delta}}\Big\rceil,
\end{align*}
where the minimization is taken over all compatibility partitionings of $\mathcal{C}$ as in Definition \ref{def:compatible cover}.
\end{theorem}

The proof of the theorem is provided in the next subsection. 

\begin{remark}
If $\mathcal{C}$ is a compatible concept class, then the sample complexity bound in Theorem \ref{thm:QERM} simplifies to $\Big\lceil{\frac{2}{\epsilon^2}\log\frac{|\mathcal{C}|}{\delta}}\Big\rceil$.
\end{remark}


\subsection{QERM algorithm and the Proof of the main result}\label{subsec:QERM}
We prove Theorem \ref{thm:QERM} by proposing our QERM algorithm. \addvg{ As in the classical ERM, our algorithm is implemented by measuring the empirical loss for each predictor $\mathcal{M}\in \mathcal{C}$ and finding the one with the minimum empirical loss. This is done by applying an appropriately designed quantum measurement on the samples to output the empirical loss value of each $\mathcal{M}\in \mathcal{C}$.} 
 In what follows, we describe this process. Further, we propose a concentration analysis for quantum measurements. 

We start with the measurement process for computing the empirical loss of only one predictor. Let $\ell: \mathcal{Y}\times \mathcal{Y} \mapsto [0,1]$ be the loss function and $\mathcal{Z}$ be the image set of $\ell$. \addvg{Since $\mathcal{Y}$ is a finite set, then so is $\mathcal{Z}$.} With that, the loss value observable for any predictor $\mathcal{M}:=\set{M_{\hat{y}}: \hat{y}\in \mathcal{Y}}$ is given by
$\MLoss:=\set{L^M_z: z\in\mathcal{Z}}$, where  
\begin{align}\label{eq:Loss operators}
 L^M_z = \sum_{\substack{y, \hat{y}\in \mathcal{Y}: \ell(y, \hat{y})=z}}  M_{\hat{y}}\tensor \ketbra{y}, \qquad  \forall z\in\mathcal{Z}.
 \end{align} 
\addvg{Therefore, the loss of $\mathcal{M}$ for predicting $y$ from a given $\rho_x$ is obtained by applying  $\MLoss$ on $\rho_x\tensor \ketbra{y}$. The result is a random variable $Z=\ell(y, \hat{Y})$ taking values from $\mathcal{Z}$ as in \eqref{eq:Loss operators}. Note that, unlike the classical settings, when the predictor and the samples are fixed the loss value is still a random variable.} In that case, the ``conditional" expectation of the loss variable $Z$ for a fixed sample is given by $\<\MLoss\>_{\rho_x\tensor \ketbra{y}}$, where $\<\cdot\>$ is the \textit{expectation value} of an observable in a quantum state. Hence, the overall expectation of $Z$ equals $\EE[Z]=\<\MLoss\>_{\rho_{XY}},$ where $\rho_{XY}$ is the average density operator of the sample. Further, it is not difficult to see that the true risk of a predictor $\mathcal{M}$ equals to  
\begin{align*}
L_D(\mathcal{M}) = \<\MLoss\>_{\rho_{XY}} =\EE[Z]=\sum_{z\in \mathcal{Z}} z \tr{L^M_z \rho_{XY}}.
\end{align*}

We compute an empirical loss of $\mathcal{M}$ by applying  $\MLoss$ on each sample. Let $z(i)$ be the realization of the loss value measured on the $i$th sample. Then, the empirical loss is given by $L_{\hat{D}}(\mathcal{M})\deq \frac{1}{n}\sum_{i}z(i).$
 Next, we provide a quantum sample complexity analysis. For that, we  present  a quantum analogous of Chernoff-Hoeffding inequality. 
\addvd{
\begin{lem}\label{lem:quantum chernoff}
Let $\rho_{i}, i\in [n]$ be \ac{iid} random density operators on a finite dimensional Hilbert space $H$. Let $\bar{\rho}=\EE[\rho_i]$ be their average density operator.  
 Let $\mathcal{M}$ be a (discrete) observable on $H$ with outcomes  bounded by the interval $[a,b]$, where $a,b\in \RR$. 
 If $V_i$ is the  outcome of $\mathcal{M}$ for measuring $\rho_i$, then for any $t\geq  0$
\vspace{-6pt}
\begin{align*}
\prob{\abs\Big{\frac{1}{n}\sum_{i=1}^n V_i - \<\mathcal{M}\>_{\bar{\rho}}]} \geq t }\leq 2 \exp{-\frac{nt^2}{2(b-a)^2}}, 
\end{align*}
   where $\<\mathcal{M}\>_{\bar{\rho}}$ is the expectation value of $\mathcal{M}$ in state $\bar{\rho}$.  
\end{lem}}
The proof is omitted as it is a direct consequence of Theorem A.19 in \cite{Ahlswede2002}.

\addvd{We apply Lemma \ref{lem:quantum chernoff} where the measurement is $\MLoss$ and the random states are our \ac{iid} samples with $\bar{\rho}=\rho_{XY}$ as the average density operator.}
Hence, by an appropriate choice of $t$,   given $\delta\in [0,1]$, with probability $(1 - \delta)$ the following inequality holds
\vspace{-10pt}
\begin{align*}
|L_{\hat{D}}(\mathcal{M})-L_{{D}}(\mathcal{M})|\leq \sqrt{\frac{2}{n}\log\frac{2}{\delta}}.
\end{align*}

As a next step, we would like to measure the empirical loss for all the predictors in the given hypothesis class. However, this is not straightforward as in the classical setting. Because,  after measuring the empirical loss of one predictor, the quantum state of the samples collapses, and we might not be able to ``reuse" the  samples to measure the loss of another predictor. Further, the no-cloning principle prohibits creating multiple copies of the training samples.

\addvf{\noindent{\textbf{Naive Strategy:}} In this strategy,  the training samples are partitioned into several batches, one for each predictor $\mathcal{M}\in \mathcal{C}$. Then, the empirical risk of each $\mathcal{M}$ is computed on the corresponding partition.} Therefore, it is easy to verify that 
\begin{align*}
\sup_{\mathcal{M}\in\mathcal{C}} |L_{\hat{D}}(\mathcal{M})-L_{{D}}(\mathcal{M})|\leq \sqrt{\frac{2|\mathcal{C}|}{n} \log\frac{2}{\delta}}.
\end{align*}
\addvf{Hence, the sample complexity of the naive strategy is $O(\frac{|\mathcal{C}|}{\epsilon^2} \log \frac{1}{\delta})$ that  blows up with the size of the hypothesis class. }

We improve upon this bound by leveraging the compatibility notion.  



\noindent\textbf{QERM for Compatible Classes:}
 Suppose the predictors in the hypothesis class $\mathcal{C}$ are compatible. Let index the elements of $\mathcal{C}$ by $\mathcal{J}=\set{1,2,..., |\mathcal{C}|}$. For each measurement $\mathcal{M}$, we have the loss observable $\MLoss$ with operators as in \eqref{eq:Loss operators}. Since $\mathcal{M}\in \mathcal{C}$ are compatible, then so are $\MLoss$. Hence, we create the POVM  $\QERMPOVM:=\set{L_{\bfz}: \bfz\in \mathcal{Z}^{|\mathcal{C}|}}$, with operators 
\begin{align}\label{eq: QERM opt}
\QERMPOVM:=\Big\{ L_{\bfz} = \prod_{j\in \mathcal{J}_\mathcal{C}} L^{M_j}_{z_j}:  \bfz \in \mathcal{Z}^{|\mathcal{C}|}\Big\},
\end{align}
where $\big\{L^{M_j}_{z}: z\in \mathcal{Z}\big\}$ are the operators of the  $\mathcal{L}_{M_j}$. 
 
 We compute the empirical loss of all predictors in $\mathcal{C}$ by applying $\QERMPOVM$ on each sample. Let $\bfz(i)$ be the outcome of  $\QERMPOVM$ when measuring the $i$th sample. By $z_j(i)$ denote the $j$th coordinate of the vector $\bfz(i)$. Then, the empirical loss of the $j$th predictor in $\mathcal{C}$ is given by 
\begin{align}\label{eq:empirircal loss compatible}
 \mathcal{L}_{\hat{D}}(\mathcal{M}_j) = \frac{1}{n}\sum_{i=1}^n z_j(i).
 \end{align} 
 Hence, we can simultaneously measure the empirical loss of all the predictors without the need for partitioning the training samples. We then establish the following result on the accuracy of the empirical loss. 
 \begin{lem}\label{lem:empirical loss compatible}
Let $\mathcal{C}$ be a finite hypothesis class consisting of compatible predictors. Let $\mathcal{L}_{\hat{D}}(\mathcal{M}_j)$ be the empirical loss of the $j$th predictor of $\mathcal{C}$ as in \eqref{eq:empirircal loss compatible}. Then, for $\delta\in [0,1]$, with probability at least $(1-\delta)$, the following inequality holds
\begin{align*}
\max_{\mathcal{M}\in\mathcal{C}} |L_{\hat{D}}(\mathcal{M})-L_{{D}}(\mathcal{M})|\leq \sqrt{\frac{2}{n} \log\frac{2|\mathcal{C}|}{\delta}}.
\end{align*}
\end{lem}
As a result, we expect that the sample complexity increases at most logarithmic with the size of the hypothesis class. Hence, we get a significant improvement over the naive strategy.


\noindent\textbf{QERM for General Classes:}
Now we extend our approach for a general hypothesis class $\mathcal{C}$.  The idea is to partition $\mathcal{C}$ into compatible subclasses as in Definition \ref{def:compatible cover}. 

\textbf{Class partitioning:} Based on Definition \ref{def:compatible}, we can check if two measurements are compatible by checking whether their operators commute. Hence, with an exhaustive search one can find all possible ways of partitioning $\mathcal{C}$ into compatible subclasses.  Note that the compatibility depends only on $\mathcal{C}$ and is independent of the samples. Hence, the partitioning can be done once as a pre-processing step. 

\textbf{Sample partitioning:} With a partitioning, observables inside each subclass can be measured simultaneously. However, each compatible class must be supplied with an exclusive set of training samples. This is because measurements belonging to different subclasses may not be compatible. In other words, the $n$ training samples have to be partitioned into multiple subsets, one for each subclass. The sample subsets are allowed to have different sizes. Let $n_j$ be the size of the $j$th subset corresponding to $j$th subclass. 

We repeat the process described in the previous part on each subclass with its sample subset. For that, we create measurements $\mathcal{L}^{\mathcal{C}_r}_{QERM}$ as in \eqref{eq: QERM opt} and compute the empirical loss of the predictors inside each subclass. We will show how to chose the batch sizes and the best partitioning of $C$. With this approach, we formally propose the QERM algorithm as presented in Algorithm \ref{alg: QERM} and establish our theorem. 

%

\vspace{-6pt}
\begin{algorithm}[h]
\caption{QERM}
\label{alg: QERM}
\DontPrintSemicolon
\KwIn{Concept class $\mathcal{C}$ and $n$ training samples.}
\KwOut{Index of the selected predictor in $\mathcal{C}$}

 Partition $\mathcal{C}$ into a set of compatible subclasses $\mathcal{C}_1, \mathcal{C}_2, ..., \mathcal{C}_m$.\;
 Partition the samples into $m$ bathes, one for each subclass. \;
 
    \For{$r=1$ \KwTo $m$}{
   Construct $\mathcal{L}^{\mathcal{C}_r}_{QERM}$  as in \eqref{eq: QERM opt} and apply it on each sample in the $r$th batch.\;
   Let $\bfz^r(i)$ be the vector outcome on the $i$th sample of batch $r$.\;
   Compute $\bar{z}^r_j=\frac{1}{n_r}\sum_i \bfz^r_j(i)$, as the empirical loss of the $j$th predictor in $\mathcal{C}_r$.\;
   } 
\Return $\argmin_{r,j} \bar{\bfz}^r_j$ as the index of the selected predictor denoted by $\mathcal{M}_{r,j}$.
\end{algorithm}
\vspace{-6pt}

As the last step in the proof of Theorem \ref{thm:QERM}, we analyze the sample complexity and find an upper bound on $n(\delta, \epsilon)$. The argument follows from standard steps.

We apply Lemma \ref{lem:empirical loss compatible} on each subclass $\mathcal{C}_r$ with the $r$th sample batch with $n_r$ samples. Set $n_j = \lceil{\frac{8}{\epsilon^2}\log\frac{2|\mathcal{C}_r|}{\delta}}\rceil$. As a result, with probability $(1-\delta)$,  the inequality 
$\max_{\mathcal{M}\in\mathcal{C}_r} |L_{\hat{D}}(\mathcal{M})-L_{{D}}(\mathcal{M})|\leq \frac{\epsilon}{2}$ holds.
Hence, from the union bound, with probability at $(1-(1-\delta)^m)\approx 1-m\delta$, we have that
$\max_{1\leq r\leq m}\max_{\mathcal{M}\in\mathcal{C}_r} |L_{\hat{D}}(\mathcal{M})-L_{{D}}(\mathcal{M})|\leq \frac{\epsilon}{2}.$
Let $\widehat{\mathcal{M}}$ and $\mathcal{M}^*$ be the predictors minimizing the empirical loss and the true loss, respectively.  Then, 
\begin{align*}
L_{{D}}(\widehat{\mathcal{M}}) \leq L_{\hat{D}}(\widehat{\mathcal{M}})+\frac{\epsilon}{2}\leq  L_{\hat{D}}({\mathcal{M}}^*)+\frac{\epsilon}{2} \leq L_{{D}}({\mathcal{M}}^*) +\epsilon.
\end{align*}
The left-hand side is the loss of the selected predictor by QERM, and the right-hand side equals $\opt+\epsilon$. Hence, the proof is complete by replacing $\delta$ with $\delta/m$.

\addvf{
\section{Conclusion}
We studied learning from quantum data and formulated the quantum counterpart of PAC framework. Then, we proposed measurement  partitioning to address the challenges such as the no-cloning principle and measurement incompatibility. Based on that, we introduce a quantum risk minimizer algorithm using which we proved bounds on the quantum sample complexity of finite concept classes.}

\section*{Acknowledgement}

This work was supported in part by
NSF Center on Science of Information
Grants CCF-0939370 and NSF Grants CCF-1524312, CCF-2006440, CCF-2007238, and Google Research Award.
\appendices

\bibliographystyle{IEEEtran}
\bibliography{main}
\end{document}